\documentclass[aps,prl,preprint,showpacs,superscriptaddress,amssymb,amsfonts]{revtex4}

\usepackage[applemac]{inputenc}

  \usepackage{epsfig}

\usepackage{graphicx}

\begin{document}
    
\title{Balanced homodyne detection in second-harmonic generation microscopy }
\author{L. Le Xuan}
\affiliation{Laboratoire de Photonique Quantique et Mol\'eculaire,
ENS Cachan, UMR CNRS 8537 , 61 Avenue du Pr\'esident Wilson, 94235 Cachan Cedex, France}
\email{roch@physique.ens-cachan.fr}
\author{F.~Marquier}
\affiliation{Laboratoire EM2C, UPR 288 CNRS, Ecole Centrale Paris,
Grande Voie des Vignes, 92295 Ch\^atenay-Malabry Cedex, France}
\author{D.~Chauvat}
\affiliation{Laboratoire de Photonique Quantique et Mol\'eculaire,
ENS Cachan, UMR CNRS 8537 , 61 Avenue du Pr\'esident Wilson, 94235 Cachan Cedex, France}
\affiliation{Laboratoire de Physique des Lasers, UMR CNRS 6627, Universit\'e Rennes 1, 35042 Rennes Cedex, France}
\author{S.~Brasselet}
\affiliation{Laboratoire de Photonique Quantique et Mol\'eculaire,
ENS Cachan, UMR CNRS 8537 , 61 Avenue du Pr\'esident Wilson, 94235 Cachan Cedex, France}
\author{F.~Treussart}
\affiliation{Laboratoire de Photonique Quantique et Mol\'eculaire,
ENS Cachan, UMR CNRS 8537 , 61 Avenue du Pr\'esident Wilson, 94235 Cachan Cedex, France}
\author{S.~Perruchas}
\affiliation{Laboratoire de Physique de la Mati\`ere Condens\'ee,
Ecole Polytechnique, UMR CNRS 7643, 91128 Palaiseau Cedex, France}
\author{C.~Tard}
\affiliation{Laboratoire de Physique de la Mati\`ere Condens\'ee,
Ecole Polytechnique, UMR CNRS 7643, 91128 Palaiseau Cedex, France}
\author{T.~Gacoin}
\affiliation{Laboratoire de Physique de la Mati\`ere Condens\'ee,
Ecole Polytechnique, UMR CNRS 7643, 91128 Palaiseau Cedex, France}
\author{J.-F.~Roch}
\affiliation{Laboratoire de Photonique Quantique et Mol\'eculaire,
ENS Cachan, UMR CNRS 8537 , 61 Avenue du Pr\'esident Wilson, 94235 Cachan Cedex, France}
\date{\today}

\begin{abstract}
We demonstrate the association of two-photon nonlinear microscopy with balanced homodyne detection for investigating second harmonic radiation properties at nanoscale dimensions. Variation of the relative phase between second-harmonic and fundamental beams is retrieved, as a function of the absolute orientation of the nonlinear emitters. Sensitivity down to $2 \sigma \approx$ 3.2 photon/s in the spatio-temporal mode of the local oscillator is obtained. This value is high enough to efficiently detect the coherent second-harmonic emission from a single $\rm{KTiOPO_{4}}$ crystal of sub-wavelength size.
\end{abstract}


\maketitle
Compared to two-photon absorption microscopy \cite{Denk_90}, non-resonant second-harmonic generation (SHG) can circumvent photobleaching issues \cite{Patterson_00}.
However SHG microscopy usually leads to weak photon flux, and extremely faint signals are therefore expected from nano-objects with second-harmonic nonlinear response. Direct detection with avalanche photodiode in photon-counting regime is a practical solution. However their sensitivity is limited by dark count noise, usually $\approx$ 10 photons per second. Fortunately, SHG is a coherent process thus interferometric detection schemes are possible, most interestingly coherent balanced homodyne detection which exhibits a sensitivity to extremely low photon flux rates \cite{Yuen_83}. Although high sensitivity with such a technique has been demonstrated to detect surface SHG emission from an air-semiconductor interface \cite{Chen_98}, to the best of our knowledge it has never been applied to the study of nano-objects. 
Since SHG phase is directly related to the sign of the associated nonlinear susceptibility, retrieving its value provides important information about the nonlinear material \cite{Stolle_96}. At nanometric scale, it allows to determine the absolute orientation of dipoles with nonlinear hyperpolarisabilities such as organic nanocrystals \cite{Brasselet_04}, to detect the polarity in cell membranes  \cite{Moreaux_00}, or to study the boundary between microscopic crystalline domains \cite{Laurell_92}. Such application prospects strongly motivate the development of phase-sensitive SHG microscopy with highly sensitive detection compared to SHG interferometric schemes previously developed  \cite{Stolle_96}. In this work coherent balanced homodyne detection is associated to SHG microscopy, allowing us to detect nano-objects with high signal-to-noise ratio and phase sensitivity.

The principle of the experiment is shown in Fig.~1. Femtosecond infrared light pulses are injected in an inverted optical microscope and tightly focused. Backward second-harmonic emission by a nonlinear crystal placed on a glass cover slide at the microscope focus is collected by the microscope objective and transmitted through a dichroic mirror toward the detection set-up. A small fraction of the fundamental beam is also reflected from the cover-slide and follows the same optical path, providing the phase reference. Second-harmonic emission of the crystal, further referred as ``signal'', is associated with an electrical field $E^{\left( 2 \omega \right)} =  \vert {{E^{\left( 2 \omega \right)}}} \vert \exp [ i \phi_{\rm{obj}}^{\left(  2 \omega \right)} ]$ , where $ \phi_{\rm{obj}}^{\left(  2 \omega \right)} $  is the phase shift of the SHG field emitted by the object as compared to the one of the incident fundamental field  $E_{\rm{in}}^{\left( \omega \right) }$ on the object. For a macroscopic nonlinear crystal, the SHG field is  $E^{\left( 2 \omega \right)} \propto \chi^{\left( 2 \right)} {E_{\rm{in}}^{\left( \omega \right) } }^{2}$ . Thus, $ \phi_{\rm{obj}}^{\left(  2 \omega \right)} $ reflects the phase of the nonlinear susceptibility $\chi^{\left( 2  \right)}$. For a nonlinear crystal in its spectral transparence window, $\chi^{\left( 2  \right)}$  is real with positive or negative values depending on the absolute orientation of the nonlinear response in respect to the crystal axis. Therefore determination of  $ \phi_{\rm{obj}}^{\left(  2 \omega \right)} $ allows to infer the absolute orientation of the nonlinear crystal.

   For the coherent optical homodyne detection of the SHG signal, the local oscillator (LO) is generated by sending part of the incident infrared beam into a BBO nonlinear crystal (Fig.~1). A Glan prism ensures linear polarization of the local oscillator along the $x$-axis and $x$-polarized signal and local oscillator are recombined by a non-polarizing 50/50 beamsplitter cube. The 180° out-of-phase interferences at the two output ports of the beamsplitter are detected by photodetectors $D_{1}$ and $D_{2}$. In balanced detection mode, the two resulting photoelectric signals $S_{1}$ and $S_{2}$ are subtracted \cite{Note}, thus canceling out their DC component. The interferometric signal is then :
\begin{equation}
      	\Delta S^{\left( 2 \omega \right)} = K \times 2 V \sqrt{ P_{\rm{LO}}^{\left( 2 \omega \right)} P_{\rm{sig}}^{\left( 2 \omega \right)}} 
	\cos\left( \phi_{\rm{sig}}^{\left(  2 \omega \right)} - \phi_{\rm{LO}}^{\left(  2 \omega \right)} \right)        
	\label{eqn1}
\end{equation}
where $P_{\rm{LO}}^{\left( 2 \omega \right)}$ and $P_{\rm{sig}}^{\left( 2 \omega \right)}$ denote the mean optical power of local oscillator and signal, $V$ is the fringe visibility, and $K$ includes the quantum efficiency of the photodiodes. The phase shift of the signal at frequency $2 \omega$ , $\phi_{\rm{sig}}^{\left(  2 \omega \right)} = \phi_{1}^{\left(  2 \omega \right)}+\phi_{2}^{\left(  2 \omega \right)}$  contains the phase shift $\phi_{1}^{\left(  2 \omega \right)}$ of the $x$-polarized radiation emitted by the nonlinear object and the phase shift  $\phi_{2}^{\left(  2 \omega \right)}$  due to propagation in the interferometer.

   In order to measure the phase shift $\phi_{\rm{sig}}^{\left(  2 \omega \right)}$, temporal interference fringes are created by displacing a mirror mounted on a piezoelectric transducer (PZT) driven with a triangular shape voltage. The resulting fringes are shown in inset of Fig.~1. To extract the phase shift information, fringes at $2 \omega$ are compared to reference fringes resulting from fundamental beam. More precisely, we perform with detectors $D_{3}$ and $D_{4}$ a separate balanced homodyne detection of the residual fundamental beam reflected by the sample by mixing it with the residual light at $\omega$ which follows the reference arm. A reference interference signal (see inset of Fig. 1) is then obtained, equal to $ \Delta S^{\left(\omega \right)} = K' \times 2 V' \sqrt{ P_{\rm{LO}}^{\left(\omega \right)} P_{\rm{sig}}^{\left(\omega \right)}} 
\cos\left( \phi_{\rm{sig}}^{\left(\omega \right)} - \phi_{\rm{LO}}^{\left(\omega \right)} \right) $  with same notations meaning as in Eq. (1). 
The quantity to be extracted is then $\Delta \phi = \left( \phi_{\rm{sig}}^{\left(  2 \omega \right)} - \phi_{\rm{LO}}^{\left(  2 \omega \right)} \right) - \left( \phi_{\rm{sig}}^{\left(\omega \right)} - \phi_{\rm{LO}}^{\left(\omega \right)}\right) $  which can be written as  $\Delta \phi =  \phi_{1}^{\left(  2 \omega \right)} + \phi_{0}^{}$, where $\phi_{0}^{}$  is an ``instrumental'' phase shift. Although $\phi_{0}^{}$ can be measured with a reference crystal its value is not required if we are only concerned in variation of $\phi_{1}^{\left(  2 \omega \right)}$  while keeping  $ \phi_{0}^{}$ constant. Finally, the value of $\Delta \phi$ is extracted using a numerical procedure. 

A macroscopic $\rm{KTiOPO_{4}}$  (KTP) crystal with axes as shown in Fig.~2 is used to test the setup.
 The $Z$-axis is parallel to the  crystal input face. With a strongly focused pump beam, a SHG backward emitted beam from the interface is observed \cite{Boyd}.
 We first check in Fig.~2b that $ \Delta S^{\left(2 \omega \right)}$ associated to this surface SHG grows linearly with the fundamental incident power on the crystal $P_{\rm{in}}^{\left(\omega \right)}$, as expected from Eq.~(1). If the crystal is rotated by 180° around the microscope optical axis ($z$-axis), the nonlinear coefficient $\chi^{\left( 2 \right)}$ is transformed in $- \chi^{\left( 2 \right)}$ and a 180° SHG-phase-shift is expected. 
Results are shown in Fig.~2c. The amplitude shows a maximum of $x$-polarized SHG when the linear polarization is aligned along the $Z$-axis of the crystal ($\psi$ = 0° or 180°) which exhibits the highest nonlinear coefficient. The full line is the theoretical result assuming an $x$-polarized incident plane wave normal to the interface and a single nonlinear emitting dipole which takes into account all nonlinear coefficients of the KTP  $\chi^{\left( 2 \right)}$ tensor. It yields good agreement with experimental points. The phase-shift $\Delta \phi$ (Fig. 2d) remains constant while the nonlinear dipole has a positive projection along the $x$-axis (0° $< \psi <$ 90°). It then endures a 180°-phase-shift when it becomes negative (90° $< \psi <$ 180°). The small deviation to the constant value is attributed to drift of the $\phi_{0}$ parameter during the step-by-step rotation of the KTP crystal.

We then apply the method to the detection of SHG from nanometric size objects. Here we use inorganic KTP crystals of sub-wavelength size ($\leq$ 200 nm). 
To first locate the crystals, the sample is raster-scanned in $x$ and $y$ directions with piezoelectric translators and SHG is detected with the avalanche photodiode as shown
 in Fig.~1. A SHG signal image of such a nano-crystal is shown in Fig.~3a. A quadratic dependence of the detected SHG intensity is observed upon varying $P_{\rm{in}}^{\left( \omega \right)}$ as expected (see Fig.~3b).
After positioning the focused infrared excitation beam at the center of the detected emission spot of a single nano-crystal, we switch to the coherent balanced homodyne detection set-up. SHG interference fringes (Fig.~3d) associated to fundamental ones (Fig.~3c) are clearly visible, giving evidence for the coherence of the nano-crystal SHG emission. Focusing the input beam next to the nano-crystal, we check that no SHG is detectable.

Since the balanced homodyne detection method consists in the projection of the signal electric field on the spatio-temporal mode of the local oscillator, mode-matching between signal and LO beams determines maximal amplitude of the fringes. This can be quantified by the fringe visibility $V$ of Eq.~(1) being equal to unity for perfect mode-matching. 
With a 2~mW LO average power, the smallest SHG fringe amplitude measured with 1~s duration averaging is equivalent to a signal power $P_{\rm{sig}}^{\left(2\omega \right)} \approx \left( 7.2 \pm 5.0\right) \times 10 ^{-19} \, \rm{W}$ assuming $V=1$. The uncertainty evaluated for 95 \% confidence interval is equivalent to 3.2 detected photons/s, close to the shot noise limit. However a unity fringe visibility is practically difficult to achieve. In an auxiliary experiment using equal powers for signal and LO beams, we measure $V=0.21$, leading to an actual sensitivity of 80~photons/s. Such a reduced value for the visibility factor is attributed to imperfect mode-matching between signal and LO modes, including polarization mismatch, wave-front distortion, and frequency chirp on the SHG-emitted femtosecond pulses.

   To conclude we have demonstrated a SHG phase-sensitive microscope with coherent balanced homodyne detection, showing a sensitivity at the photon/s level. The high-spatial resolution and the sensitivity of the technique is well-adapted to the study of SHG from nano-crystals. Lock-in detection techniques, requiring higher interferometer stability, will allow us to detect much lower flux of photons.

\begin{acknowledgments}
We are very grateful to V. Le Floc’h for first observation of SHG fringes, and to 
D.~Lupinski and Ph.~Villeval at Cristal Laser S.A. (Nancy, France), for providing us with the KTP material. We acknowledge fruitful discussions with F.~Grosshans, J.~Zyss, 
J.-J.~Greffet and R.~Carminati, and the help of R.~Hierle for sample preparation. This work is supported by the AC~Nanoscience research program and by Institut Universitaire de France.
\end{acknowledgments}


\clearpage

\begin{figure}  
\centerline{\includegraphics[width=15cm]{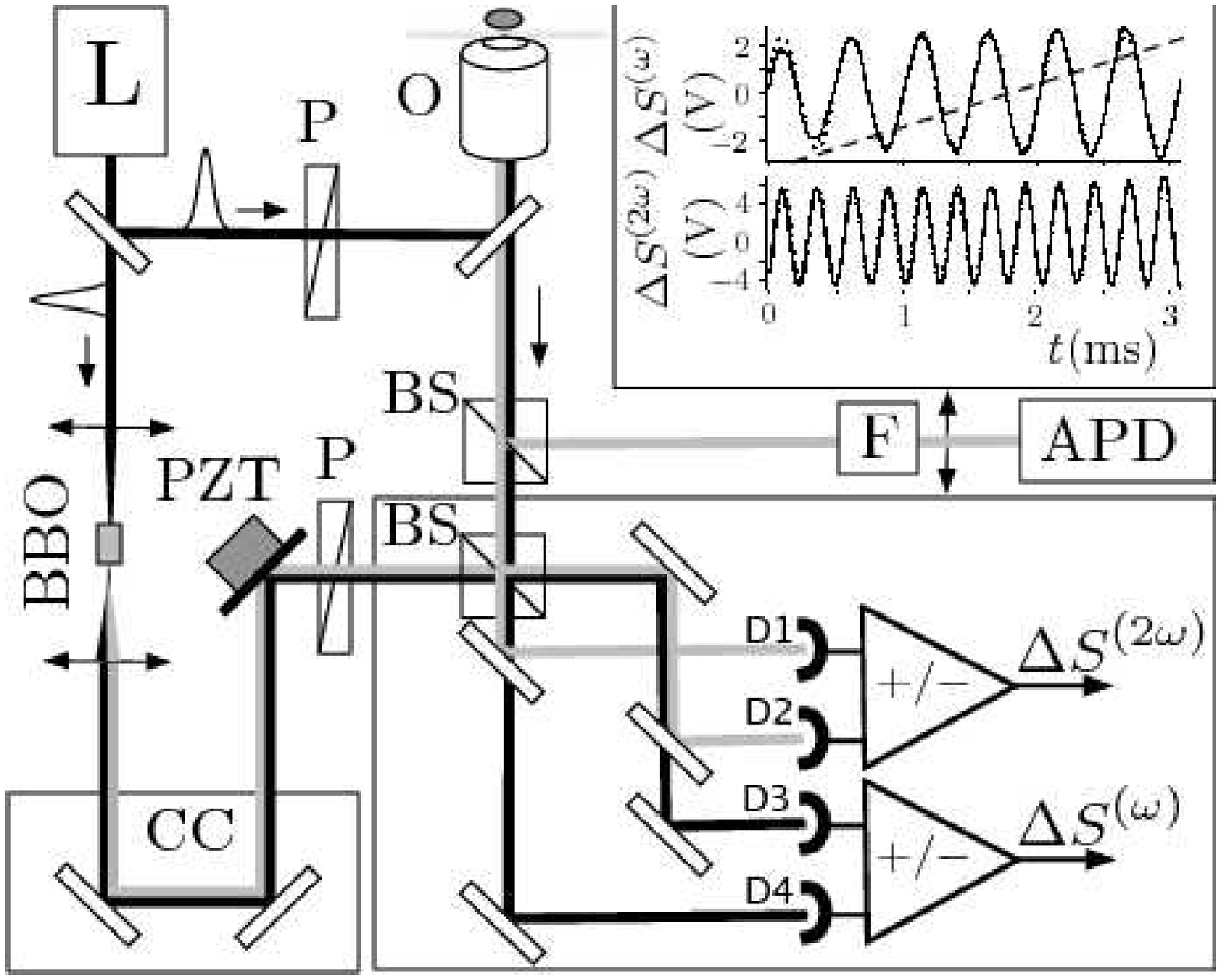}}\label{setup}
\caption{Schematic of the experimental set-up. L: femtosecond laser (86 MHz repetition rate, $\approx$ 100 fs pulse duration, $\lambda =$ 986 nm); BBO: nonlinear $\chi^{\left( 2 \right)}$ crystal; P: Glan prism; O: microscope objective (N.A. = 1.40, $\times$ 100) leading to a $\approx$ 300 nm FWHM diameter focal spot; BS: non-polarizing beamsplitter; CC: corner-cube; PZT: mirror mounted on a piezoelectric transducer; KTP: macroscopic $\rm{KTiOPO_{4}}$ crystal or single sub-wavelength size crystal; F: SHG filter; APD: avalanche photodiode in photon counting regime; $D_{1}$ and $D_{2}$: p-i-n  Si photodetectors of the balanced receiver recording SHG at $2 \omega$  (a SHG filter, not shown, is put in front of the detectors); $D_{3}$ and $D_{4}$: the same for the balanced receiver at fundamental optical frequency $\omega$ (an IR filter, not shown, is put in front of the detectors). {\it{Insert}}: full lines : detected signals as a function of time for a forward translation of the mirror. Up: signal $\Delta S^{\left( \omega \right)}$  at fundamental frequency with a sinusoidal fit at frequency $f =$ 1.9 kHz, and bottom:  $\Delta S^{\left( 2 \omega \right)}$ at SHG frequency with a sinusoidal fit at frequency $2 f$ = 3.8 kHz, dashed line:  voltage applied to the PZT on half a period (vertical scale: a.u.).}
\end{figure}

\clearpage

\begin{figure}  
\centerline{\includegraphics[width=15cm]{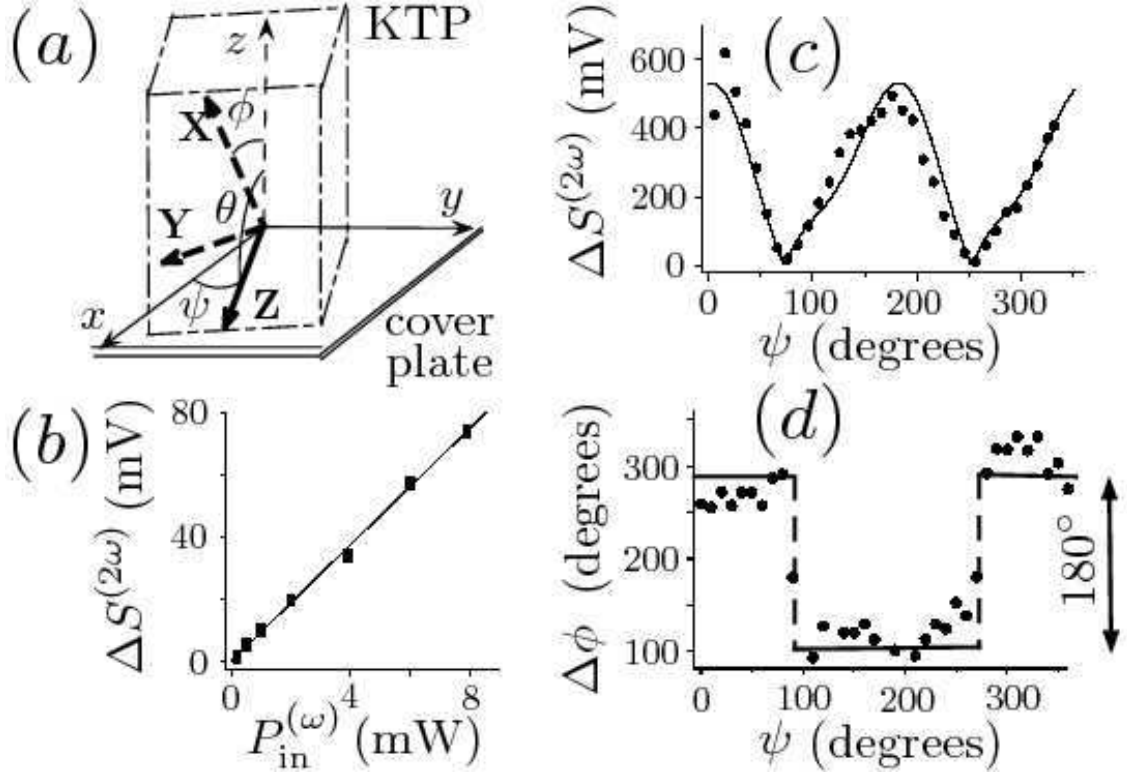}}\label{calibration}
\caption{Results from the balanced homodyne detection. (a) Laboratory axes ($x,y,z$) with $x$-polarized incident fundamental beam propagating along $z$-axis. ($\bf{X},\bf{Y},\bf{Z}$) $\rm{KTP}$  crystal principal axes ($\theta =$ 90° and $\Phi =$ 23.5°). Full lines denote axes in the ($x,y$) plane. $\psi$ : rotation angle of the $\chi^{\left( 2 \right)}$  nonlinear crystal around the $z$-axis. (b) Linearity of   $\Delta S^{\left( 2 \omega \right)}$ as a function of fundamental incident power. Points : experimental data, full line : best linear dependence fit. (c) :  $\Delta S^{\left( 2 \omega \right)}$ as a function of $\psi$ (in degrees). (d) : Relative SHG phase shift $\Delta \phi$ as a function of $\psi$.}
\end{figure}

\clearpage

\begin{figure}
\centerline{\includegraphics[width=15cm]{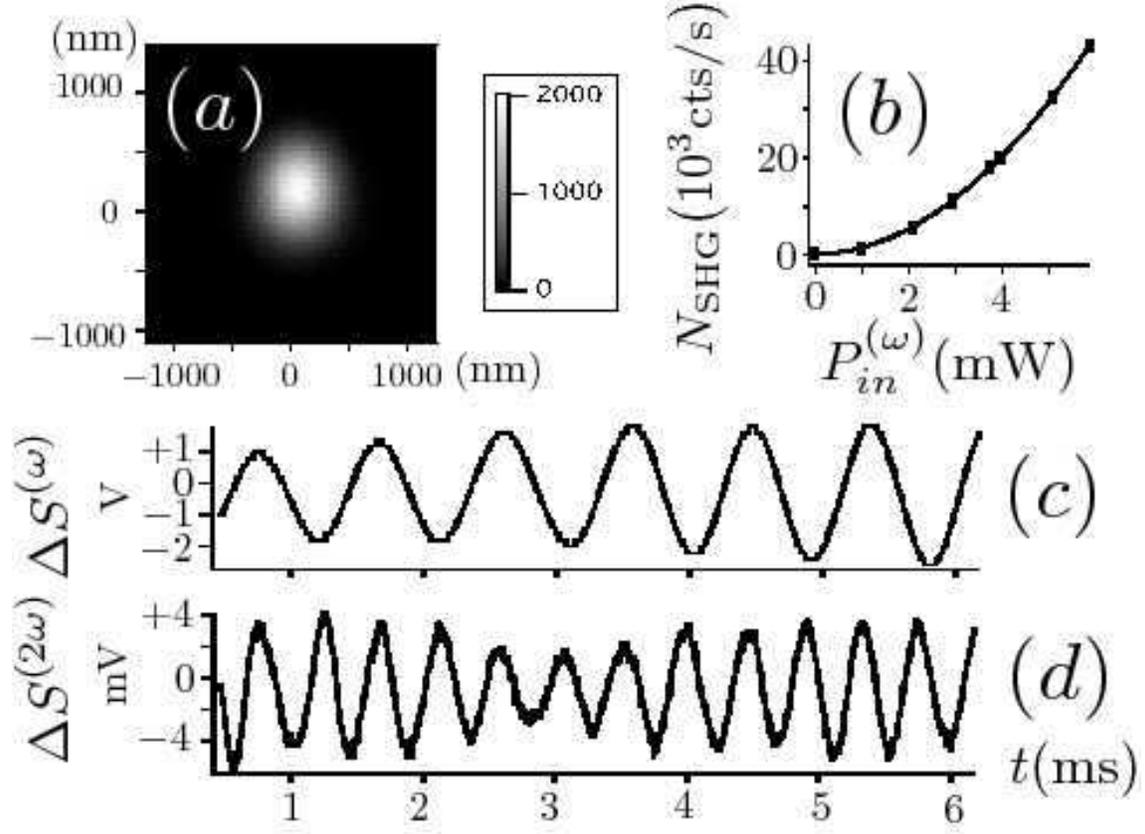}}\label{NanoKTP}
\caption{Application of the balanced homodyne technique to sub-wavelength-size nonlinear crystals. (a) Raster-scan image of SHG signal from a KTP nano-crystal with an avalanche photodiode. The FWHM diameter of SHG spot intensity is about 300 nm, very close to the theoretical two-photon microscope resolution (320~nm FWHM). (b) Number of detected SHG photons as a function of incident power with a best square-law dependence fit. (c) Fundamental interference signal $\Delta S^{\left( \omega \right)}$ as a function of time for a single nano-crystal and (d) corresponding SHG interference signal $\Delta S^{\left( 2 \omega \right)}$.}
\end{figure}

\end{document}